\RequirePackage{amsmath,amssymb,amsfonts}
\documentclass[runningheads]{llncs}
\usepackage[T1]{fontenc}
\usepackage{graphicx}
\usepackage{algorithmic}
\usepackage{multirow}

\begin{document}
\title{Physics-Guided Multi-View Graph Neural Network for Schizophrenia Classification via Structural-Functional Coupling}
\titlerunning{Physics-Guided Multi-View GNN via SC-FC Coupling}

\author{Badhan Mazumder\inst{1,2}\and
Ayush Kanyal\inst{1,2} \and
Lei Wu\inst{2}  \and
Vince D. Calhoun \inst{2} \and
Dong Hye Ye\inst{1,2}}
\authorrunning{B. Mazumder et al.}
%
\institute{Department of Computer Science, Georgia State University, Atlanta, GA, USA \and
Tri-Institutional Center for Translational Research in Neuroimaging and Data Science (TReNDS), Georgia State University, Georgia Institute of Technology, Emory University, Atlanta, GA, USA\\
\email{\{bmazumder1,akanyal1,lwu9,vcalhoun,dongye\}@gsu.edu}}
\maketitle              

\begin{abstract}
Clinical studies reveal disruptions in brain structural connectivity (SC) and functional connectivity (FC) in neuropsychiatric disorders such as schizophrenia (SZ). Traditional approaches might rely solely on SC due to limited functional data availability, hindering comprehension of cognitive and behavioral impairments in individuals with SZ by neglecting the intricate SC-FC interrelationship. To tackle the challenge, we propose a novel physics-guided deep learning framework that leverages a neural oscillation model to describe the dynamics of a collection of interconnected neural oscillators, which operate via nerve fibers dispersed across the brain's structure. Our proposed framework utilizes SC to simultaneously generate FC by learning SC-FC coupling from a system dynamics perspective. Additionally, it employs a novel multi-view graph neural network (GNN) with a joint loss to perform correlation-based SC-FC fusion and classification of individuals with SZ. Experiments conducted on a clinical dataset exhibited improved performance, demonstrating the robustness of our proposed approach.
\keywords{Structural-Functional Coupling \and Physics-guided Learning \and Graph Neural Network \and Brain Connectivity  \and Schizophrenia}
\end{abstract}

\section{Introduction}
Schizophrenia (SZ) is a complex neuropsychological condition with an unclear etiology~\cite{1}. As SZ symptoms overlap with those of other mental disorders, timely diagnosis and treatment present significant challenges. While the behavioral and cognitive aspects of SZ have been extensively studied, the intricate relationship between structural connectivity (SC) and functional connectivity (FC) in the brain has only recently garnered attention for its potential in understanding SZ causation~\cite{2}. Research suggests these relationships could serve as important biomarkers for SZ diagnosis \cite{3,4}. FC, obtained from resting-state functional magnetic resonance imaging (rs-fMRI), reflects the interactions between brain regions and has been identified as a vital biomarker for SZ \cite{L_10,L_9,L_13,L_11}. However, acquiring functional data is often time-consuming and costly. In contrast, SC, derived from diffusion tensor imaging (DTI), provides estimation of brain connectivity through structural brain fibers, and is more readily available. Investigating SC-FC coupling could not only pinpoint potential SZ biomarkers but also aid in mapping FC from SC \cite{L_5,L_6,L_4}. 

In recent years, deep learning approaches have been employed to investigate SZ from both single and multi-modal perspectives \cite{5,L_12,6}. Additionally, attempts have been made to map FC from SC using traditional deep-learning approaches to elucidate neuropsychological conditions \cite{L_2,8,7}. Existing approaches have shown a deficiency in incorporating a comprehensive understanding of neuroscience at a systemic level. In particular, most of these methods primarily focus on identifying statistical correlations between SC and FC topology patterns. However, the human brain is considered a complex system with non-random functional fluctuations, supported by coherent system-level mechanisms and frequency-specific neural oscillations \cite{L_4,L_3,L_2}. Consequently, they lack a systematic integration approach to elucidate the coupling mechanism governing neural population communication and the emergence of notable brain functions atop structural connectomes. Furthermore, limited research has explored leveraging SC-FC coupling to distinguish SZ individuals from healthy controls (HCs).

To tackle the challenges, we propose a physics-guided deep learning framework for classifying SZ by learning the interrelation between SC and FC at the system dynamics level. Initially, SC-FC coupling was leveraged by employing the Kuramoto model \cite{9}, a dynamic oscillator model used to analyze synchronized events in complex systems. Utilizing the Kuramoto model in conjunction with U-Net, FC were mapped from SC, illustrating the physical coupling of oscillatory neural units through nerve fibers to elucidate the self-organized fluctuation patterns evident in BOLD signals. Afterwards, we treated each modality as a separate view and employed a novel graph neural network (GNN) with a joint loss, including both canonical correlation analysis (CCA) loss and classification loss. In contrast to previous multi-modal approaches, we exclusively utilized SC as input, dynamically learning unique embeddings from both SC and generated FC, and then combined them to classify SZ with enhanced precision.

In a nutshell, our work makes a dual contribution in essence:
\vspace{-5pt}
\begin{itemize}
\item \textbf{Physic-guided Deep Learning for SC-FC Coupling}: We developed a novel physics-guided deep learning framework predicting FC from SC, addressing data scarcity in the functional domain by capturing SC-FC coupling dynamics effectively.
\item \textbf{Multi-view GNN for SZ Classification}: We developed a novel multi-view GNN with a joint loss, which not only correlates SC and FC but also classifies SZ individuals from HC simultaneously.
\end{itemize}

\section{Methodology}
\begin{figure}[t]
\centerline{\includegraphics[width=\textwidth ]{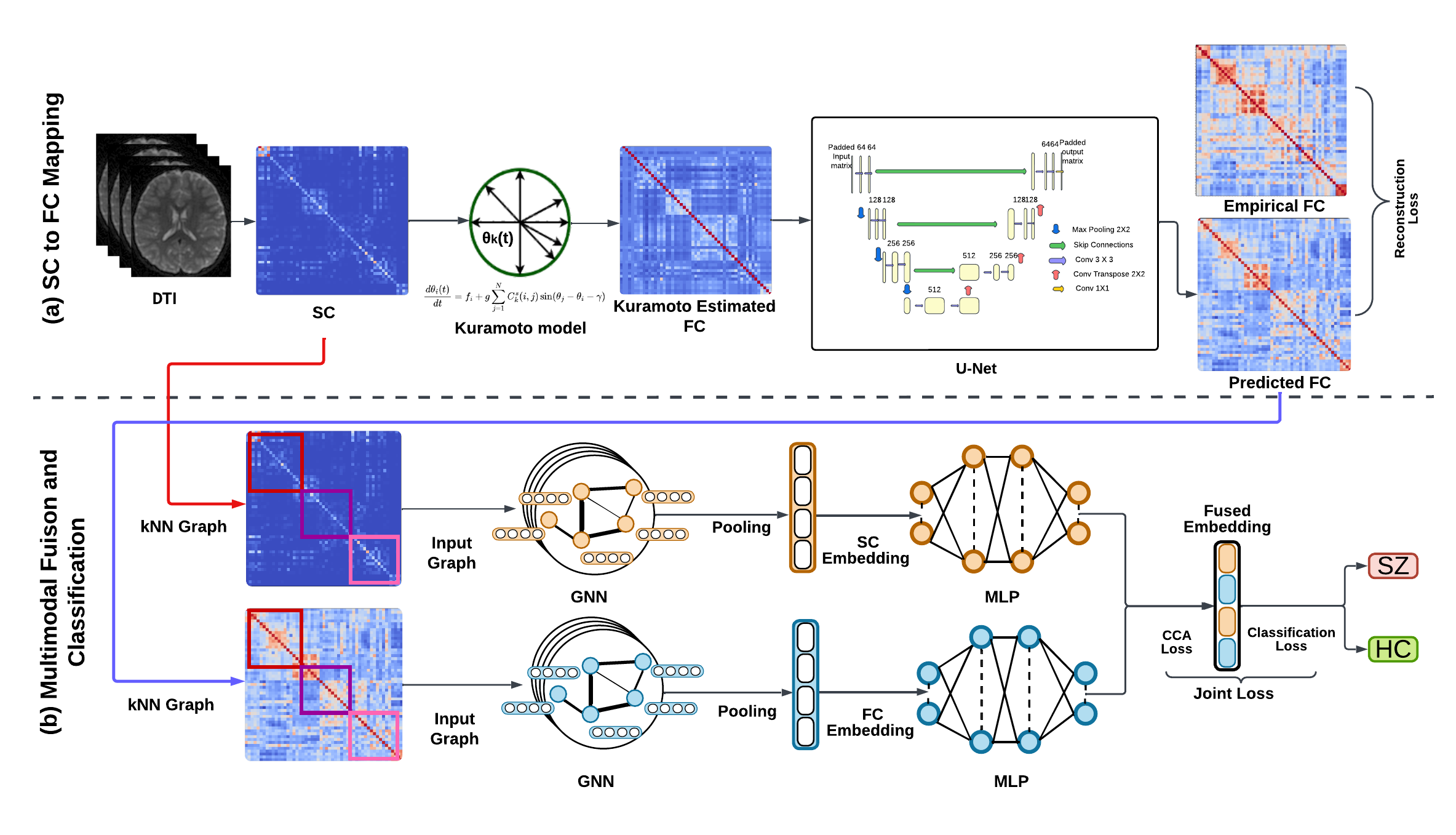}}

\caption{Overview of the proposed framework: (a) SCs were provided as input to the Kuramoto model for FCs simulation which were then input to a U-Net to predict more precise FCs. (b) Both SCs and predicted FCs were treated as separate views and fed into GNNs followed by multi-layer perceptrons (MLPs) to learn low dimensional embeddings which were used to compute a joint loss consisting both CCA-based loss and classification loss for optimizing the whole network.} \label{fig1}
\vspace{-10 pt}
\end{figure}

\vspace{-5 pt}
Suppose our provided dataset be, $ D= \left \{ \left ( C_{k}^{s}, C_{k}^{f} \right ), Y_{k} \right \}_{k=1}^{E} $ with total number of subjects $E$ where for each of the $k_{th}$ subject $C_{k}^{s}$ and $C_{k}^{f}$  represents the SC and FC, respectively, along with $Y_{k}$ as corresponding label (SZ: $1$ or HC: $0$). The goal is to predict $\hat C_{k}^{f}$ (FC) from given input $C_{k}^{s}$ (SC) and employ both of them to learn a rich correlated low-dimensional embedding for each $k_{th}$ subject and utilize that to perform binary classification. As depicted in Fig. \ref{fig1}, our proposed approach consists of two main phases. Initially, we considered $C_{k}^{s}$ as input and generated $\hat C_{k}^{f}$ by employing Kuramoto simulations followed by a U-Net. Subsequently, we integrated correlation-based fusion of multi-modal representations and SZ classification seamlessly in an end-to-end setting.

\vspace{-5 pt}
\subsection{Kuramoto Model-based SC to FC Mapping}
\subsubsection{Kuramoto Model-based FC Simulation}
Kuramoto model \cite{9} is a physics-guided mathematical model that describes the synchronization within a system of coupled oscillators with varying frequencies. Compared to  generative AI methods like GAN (Generative Adversarial Network) and VAE (Variational Autoencoder), which face several performance issues with real world complex data such as large training dataset requirement, mode collapse, limited interpretability and lack a neuroscience perspective \cite{GAN_1,GAN_2,VAE}, the Kuramoto model can function not only with limited data and leverage SC-FC coupling but also provide biological plausibility, interpretability, parameter control, computational efficiency and insights into brain dynamics to generate FCs more precisely using SCs, as showed in several recent studies \cite{L_4,L_3,L_2}.

We treated each brain region as its dynamic oscillator system and aimed at estimating the collective coherence that emerged from interactions of distinct brain regions. Mathematically, the Kuramoto model with $N$-brain regions can be described as follows:
\vspace{-5pt}
\begin{equation}
\frac{d\theta_{i}(t)}{dt} = f_{i} + g\sum_{j=1}^{N}C_k^{s}(i, j)\sin(\theta_{j} - \theta_{i} - \gamma),
 \label{kuromoto}  
 \end{equation} 
where $\theta_{i}$ denotes $i$-th brain region's phase state, $f_{i}$ is the $i$-th brain region's intrinsic frequency, $g$ represents the global coupling strength, and $\gamma$ presents the phase shift and $C_k^{s}(i, j)$ defines the coupling strength from the $i$-th to the $j$-th region for the $k$-th subject. Following this, we employed the Balloon-Windkessel hemodynamics model \cite{10,li2018analyzing} which is capable of computing the neural activity of interacting brain regions from phase information. This can be mathematically expressed as follows:
\begin{equation}
a_i(t) = a_0 \sin(\theta_i(t)),
 \label{hemodynamics}  
 \end{equation}
 where $a_0$ represents the amplitude, and $a_i(t)$ denotes the time-series neural signal from $i$-th brain region. Finally, we computed the neural activity correlations using Equ. \ref{corr} between different brain regions to generate the Kuramoto simulated FC matrices.
\begin{equation}
Corr_{p}(i, j) = \frac{\langle a_i(t) a_j(t) \rangle - \langle a_i(t) \rangle \langle a_j(t) \rangle}{std(a_i(t)) std(a_j(t))}
 \label{corr}  
 \end{equation}  
Here, $std$ denotes the standard deviation.
\vspace{-5pt}
\subsubsection{U-Net based FC Prediction}
We coupled a U-Net\cite {11} with Kuramoto model to improve the $C_{k}^{s}$ to $\hat C_{k}^{f}$ mapping. Since the Kuramoto model is simulation-based, we employed U-Net to leverage its efficient convolutions and skip connections to retain the spatial information and preserve both local-global context for performing FC prediction with more precision in a data-driven manner. Moreover, it is easy to incorporate the Kuramoto model into a U-Net for network optimization due to its simple architecture. 
    
The U-Net has a symmetric encoder-decoder structure that includes a down-sampling and up-sampling path. We utilized 4 down-sampling and up-sampling blocks in our employed U-Net. Each down-sampling block consisted of two convolutional layers with ReLU activation followed by max pooling. As we down-sampled, the spatial dimensions were reduced and the network doubled the number of feature channels. The up-sampling path mirrored this and added features from corresponding blocks of the down-sampling path utilizing skip connections. The up-sampling blocks also included two convolution layers with ReLU activation, followed by transposed convolution, which reduced the feature channels and increased the spatial dimensions. For the last layer, we used a $1 \times 1$ convolution kernel which produced a $64 \times 64$ matrix that was subsequently cropped to the original $53 \times 53$ matrix. During training, we integrated the outcome obtained from the Kuramoto model in our U-Net and jointly optimized by employing $\mathcal{L}_{U}$ as our loss function to gain a significant boost in terms of the correlation between $\hat C_{k}^{f}$ and $C_{k}^{f}$.

\vspace{-10pt}
\begin{equation}
\mathcal{L}_{U} = \frac{1}{E} \sum_{k=1}^{E} (\hat{C}_{k}^{f} - C_{k}^{f})^2 + \lambda_{w}\mathcal{F}(\hat C_{k}^{f})
\label{loss}  
\end{equation}
Here, $(\hat{C}_{k}^{f} - C_{k}^{f})^2$ term represents the classical mean square error (MSE) loss, $\mathcal{F}(\hat C_{k}^{f})=  \|  (\hat{C}_{k}^{f} | \theta_{i}(t), a_i(t), Corr_{p}(i, j)) \|_{2}^{2}$ defines regularization term from Kuramoto simulation outcome and $\lambda_{w}$ denotes the regularization parameter controlling the balance between minimizing model weights to prevent over-fitting and properly fitting the training set.   
\vspace{-5 pt}
\subsection{Multi-modal Fusion and Classification}
\subsubsection{Modality-wise Graph Formation}
A weighted graph $C_{k}^{*}=\left ( A, B_{k}^{*}, W_{k}^{*} \right)$ was utilized to delineate each modality within connectomics. Here, $ A=\left \{ a_{j} \right \}_{j=1}^{Q}$ signifies a collection of nodes, with a cardinality of $ Q $, as defined by Region Of Interests (ROIs) in connectomics. The set $ B_{k}^{*}=A \times A $ represents the edges and connections between ROIs, encapsulated by the weighted adjacency matrix $ W_{k}^{*} \in \mathbb{R}^{Q \times Q}$.
The strength of connectivity between two ROIs dictates the edge weights within $C_{k}^{*}$. To streamline the graph structure, a k-NN graph was derived from $C_{k}^{*}$ by retaining only the top $k$ neighbors based on connection strength for each node, thereby minimizing the graph's overall edge count. The parameter $k$ governs the sparsity of the graph, tailored to its topological arrangement. We sparsified the graph using k-NN to prioritize significant connections among neighbors, minimizing edge count and enhancing SZ classification performance by disregarding distant, irrelevant connections \cite{L_1}. In this work, we investigated the impact of different values of $k$ (specifically, $k=3, 5, 10,$ and $20$) where for $k=5$ best performance was observed and after that performance degradation was monitored with increasing $k$ value. To capture the local statistical characteristics of each connectomics modality, Local Degree Profiles (LDP) \cite{12} were computed from the $C_{k}^{*}$ graph and considered as node features $L_{p}$. 
\begin{equation}
L_{p} = [Deg(p)\oplus Mean(S_{p})\oplus Std(S_{p}) \oplus Min(S_{p}) \oplus Max(S_{p})],
 \label{E0}  
\end{equation}
where $S_{p}={Deg(o)|(p, o)\in B_{k}^{*}}$ presents the degree statistics within the two-hop neighborhood of node $p$, and $\oplus $ denotes concatenation.
\vspace{-10pt}
\subsubsection{GNN-based Multi-modal Fusion and Classification}
After gathering node features corresponding to each modality graph, we utilized data from both views: $C_{k}^{s}$ and $\hat C_{k}^{f}$; to employ an encoder aimed at learning data patterns. Given the widespread utilization of GCN \cite{13} and its variations in capturing both semantic and structural information within graph data, we opted for a GCN model as the encoder, to derive embedding for all subjects within each perspective. It is essential to highlight the adaptability of the graph encoder to accommodate other potent graph learning encoders such as GAT \cite{14} and GIN \cite{15}. However, for consistency and ease of comparison, we adhered to the commonly employed GCN model as the encoder in this study. The operation of GCN on the $m$-th hidden layer can be described mathematically as follows:
\begin{equation}
I^{m+1}=\sigma(D^{-1/2}B_{k}^{*}D^{-1/2}I^{m}\theta^m)
\label{E1}
\end{equation}
Here, $\sigma(\cdot)$ represents an activation function, and $D$ denotes the diagonal matrix derived from $B_{k}^{*}$. $I^{m}$ represents the output features at the $m$-th layer while $\theta^m$ denotes the trained weight matrix. To ensure comparability of embeddings, we employed a shared GCN encoder to project both views into a common embedding space and generated normalized embedding matrices for both views which were then fed into two following MLPs to obtain into lower dimensional embeddings: $X^{s}_{k}$ and $X^{f}_{k}$.

Deep models may struggle to generalize effectively on test data if they rely too heavily on irrelevant features. To address this issue, incorporating input-consistency regularization can reduce the correlation between irrelevant features, thus enhancing testing performance \cite{16}. Additionally, Canonical Correlation Analysis (CCA), which matches the similarity of two representations, is commonly utilized in tasks such as multi-view fusion \cite{17,18,19}. The fundamental objective of deep CCA can be stated as:
\begin{equation}
\begin{aligned}
max_{G} L_{CCA}= \mathbb E_{V^{s},V^{f}}[G(V^{s})^\top G(V^{f})] \\
 s.t. \quad Z_{G(V^{s}),G(V^{s})}=Z_{G(V^{f}),G(V^{f})}=\mathbf{I} 
\label{E2}
\end{aligned}
\end{equation}
Here, $G$ represents a normalized non-linear embedding: $V\rightarrow X$ and $G\leftarrow  G/||G||_{2}$ in terms of both views $V^{s}$ and $V^{f}$, $Z$ indicates the covariance matrix that $ Z_{G,G}=\mathbb E_{V^{s}} [G(V^{s})G(V^{s})]$.

Motivated by this, we employed a generalized CCA-based loss $\mathcal{L}_{CCA}$ along with cross-entropy loss $\mathcal{L}_{CE}$ for classification task as a weighted joint loss function $\mathcal{L}_{joint}$ in our work which can be expressed as follows:

\begin{equation}
\resizebox{\textwidth}{!}{$
\mathcal{L}_{CCA} = -\frac{1}{E}\sum_{k=1}^{E} G(V^{s}_k)^\top G(V^{f}_k) + \lambda_{z} \left( \|{Z}_{G(V^{s}),G(V^{s})} - \mathbf{I}\|_F^2 + \|{Z}_{G(V^{f}),G(V^{f})} - \mathbf{I}\|_F^2 \right)
$}
\label{E3}
\end{equation}
Here, $G(V^{s}_k)$ and $G(V^{f}_k)$ represents the transformed data from obtained $X^{s}_{k}$ and $X^{f}_{k}$ for $k_{th}$ subject. $\|{Z}_{G(V^{s}),G(V^{s})} - \mathbf{I}\|_F^2 + \|{Z}_{G(V^{f}),G(V^{f})} - \mathbf{I}\|_F^2 $ define the penalty terms that penalize deviation of the empirical covariance matrices of the transformed data from the identity matrix $\mathbf{I}$ and $\lambda_{z}$ is a hyperparameter that regulates the balance between maximizing the canonical correlation and ensuring the transformed representations remain decorrelated and normalized. It adjusts the weights of the mentioned penalty terms.
\begin{equation}
\mathcal{L}_{CE} = -\frac{1}{E} \sum_{k=1}^{E} \left( Y_{k} \log(\hat{Y}_{k}) + (1 - Y_{k}) \log(1 - \hat{Y}_{k}) \right)
\label{E4}
\end{equation}
Here, $\hat{Y}_{k}$ defines the obtained predicted class label.
\begin{equation}
\begin{aligned}
\mathcal{L}_{joint}= \alpha * \mathcal{L}_{CCA} + \beta * \mathcal{L}_{CE}
\label{E5}
\end{aligned}
\end{equation}
Here, $\alpha$ and $\beta$ denote the weight for the $\mathcal{L}_{CCA}$ and $\mathcal{L}_{CE}$, respectively.

After the lower dimensional embeddings $X^{s}_{k}$ and $X^{f}_{k}$ were computed for two views, the subsequent step involved utilizing the $\mathcal{L}_{joint}$ function to train and optimize the entire network. This approach aimed to enhance the correlation between embeddings $X^{s}_{k}$ and $X^{f}_{k}$ while also leveraging them for SZ classification tasks employing a linear layer seamlessly in an end-to-end fashion.
\vspace{-6 pt}
\section{Experiments}
\vspace{-5 pt}
\subsection{Dataset and Data Preprocessing}
To validate our approach, we analyzed a subset from the Function Biomedical Informatics Research Network (FBIRN) \cite{20} dataset which consisted of rs-fMRI and DTI scans from both individuals diagnosed with SZ and HC. The FBIRN study involved individuals ranging in age from 18 to 62 years, with a mean age of 37.19 years. Altogether, our employed subset included 277 participants, with 141 classified as SZ individuals and 136 as HC. 

We utilized the Neuromark \cite{21} pipeline, an automated spatially constrained method based on independent component analysis (ICA), to derive $53 \times 53$ functional network connectivity (FNC) matrices from rs-fMRI. Initially, an adaptive-ICA approach was used to estimate unique functional components for each subject along with their respective time courses. After that, the FNCs were determined by assessing correlations among the temporal patterns of 53 intrinsic connectivity networks (ICNs). For DTI, $dtifit$ (FSL) \cite{22} was employed with linear regression. Whole-brain deterministic tractography was conducted using diffusion tensor data and track (CAMINO) \cite{23} and streamlines were obtained in native space during tractography. We normalized to the same NeuroMark atlas \cite{21}, then mapped the native fractional anisotropy (FA) map to standard MNI space using ANTs \cite{24}. Finally, $53 \times 53$ SC matrices were computed, indicating fiber counts between regions of interest (ROIs) within the NeuroMark atlas by tallying streamlines passing through each ROI pair.

\vspace{-10 pt}
\subsection{Implementation Details}
 
In Kuramoto model, the initial phases were initialized randomly within $0$ to $2\pi$. The simulation period, global coupling strength ($g$), and phase shift ($\gamma$) were set to 30.597, 7.5929 and 0.56713, respectively. In Eq. \ref{hemodynamics}, we defined the baseline amplitude ($a_0$) of the neural activity signals as 10. All of our experiments followed a 5-fold cross-validation setup, dividing data into 80\% for training and 20\% for testing to ensure robust evaluation of our method. For the U-Net, during training, the learning rate was kept at 0.001. Our employed GCN had 5 layers with a channel size of 32. Moreover, ReLU was employed as the activation function, ADAM (Adaptive Moment Estimation) served as the optimizer with a learning rate of 0.001, dropout value of 0.2, and batch size of 64. The end-to-end network was trained up to 500 epochs. As for assessing the classification performance, we computed accuracy, precision, and f1-score as evaluation metrics. 
\begin{table}[t]
\caption{Comparison of classification score with baselines for 5-fold cross validation [Unit: \%; (Mean $\pm$ Standard Deviation)].}\label{tab3}
\centering
\begin{tabular}{|c|c|c|c|c|}
\hline

Method & Modality & Accuracy & Precision & f1-score \\ \hline
GCN & SC & 68.48 $\pm $11.49  & 69.06 $\pm$ 12.68 & 66.29 $\pm$ 11.30 \\ 
GAT \cite{14} & SC + FC & 68.64 $\pm$ 6.17 & 70.16 $\pm$ 5.42 & 64.65 $\pm$ 6.63\\ 
GCNN \cite{GCNN}& SC + FC & 68.11 $\pm$ 3.05 & 68.61 $\pm$ 3.25 & 67.88 $\pm$ 3.09 \\ 
BrainNN \cite{BrainNN}& SC + FC & 72.56 $\pm$ 2.08 & 72.30 $\pm$ 2.05 & 72.23 $\pm$ 8.16\\
M-GCN \cite{MGCN}& SC + FC & 75.35 $\pm$ 3.72& 75.80 $\pm$ 3.58 & 74.62 $\pm$ 3.62\\ 
\textbf{Proposed} &\textbf{SC + FC} & \textbf{78.68 $\pm$ 4.31} & \textbf{78.89 $\pm$ 4.49} &  \textbf{78.58 $\pm$ 4.35} \\ 
\hline
\end{tabular}
\vspace{-10 pt}
\label{Table3}
\end{table}

\subsection{Quantitative Evaluation}
\vspace{-5 pt}

The main novelty of our proposed method lies in utilizing SCs to simultaneously generate FCs by learning SC-FC coupling for SZ classification. Unlike existing methods focused solely on generation or classification, our framework simultaneously performs both tasks end-to-end. To the best of our knowledge, we are the first ones to introduce such an approach in this domain. Our model generates FCs (symmetrical around the diagonal like fMRI-derived ones) solely utilizing SCs as inputs (obtained from DTI) and integrates them for SZ classification, eliminating the need for fMRI data in the process. 

Initially, our predicted FCs using only Kuramoto model had a correlation of $0.561 \pm 0.042$ (Mean $\pm$ Standard Deviation) across 5-folds with the empirical FCs. For comparison purpose, we also utilized only U-Net where our obtained correlation was $0.425 \pm 0.215$. However, an improved correlation of $0.748 \pm 0.012$ was obtained when we coupled both Kuramoto and U-Net together as proposed which directly validates our hypothesis of employing U-Net with Kuramoto in the first place. 

For proper validation, we compared our proposed approach by considering five existing state-of-the-art (SOTA) deep learning methods as baselines and reported the obtained outcomes in Table \ref{Table3}. Among these methods; GCNN \cite{GCNN}, BrainNN \cite{BrainNN} and M-GCN \cite{MGCN} are tailored for GNN based multimodal fusion and end-to-end classification. To mention, these baselines require both structural and functional modalities as input to the model for producing the final classification outcome, whereas our proposed method only requires the structural modality.

In Table \ref{Table3}, it is evident that our proposed method demonstrates superior performance across all three indices. Specifically, our approach surpassed previous SOTA GNN based multimodal fusion baselines: GCNN \cite{GCNN}, BrainNN \cite{BrainNN}, and M-GCN \cite{MGCN}, with improvements of up to 10\% across all three indices. This not only highlights the superiority of our proposed approach but also the effectiveness of our proposed CCA-based multimodal fusion of graph representations. Additionally, we also employed GCN on SC data solely to compare our proposed approach from single modality view as reported in Table \ref{Table3} where we obtained an improvement of 10.20\%, 9.83\% and 12.29\% in terms of in accuracy, precision and f1-score, respectively; which validates our hypothesis of utilizing multimodal connectomics data to learn SC-FC coupling for improved SZ classification. 

\begin{figure}[t]
\centering
\includegraphics[width= \textwidth, height= 9 cm]{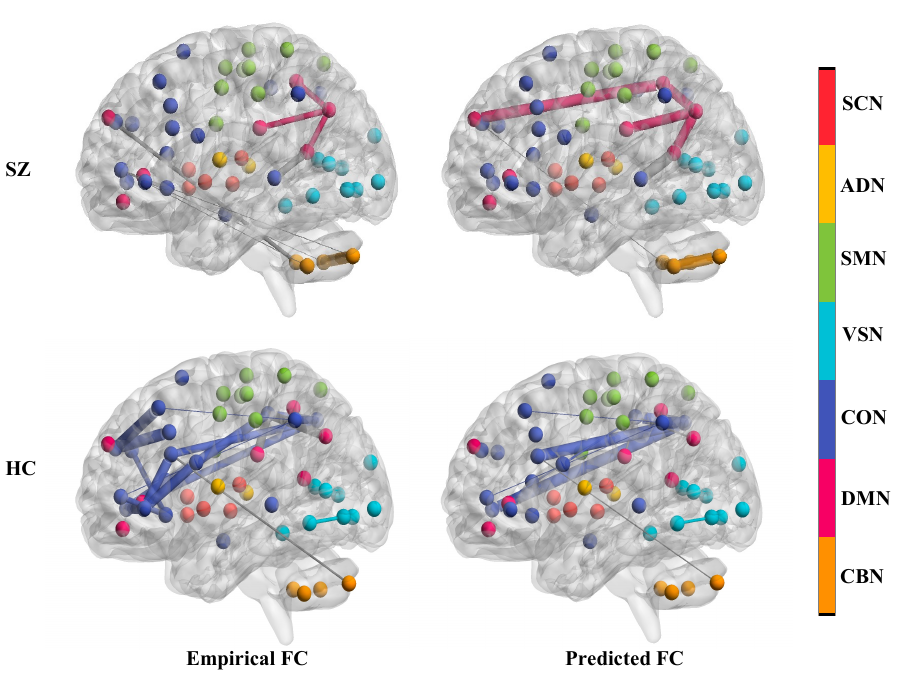}
\caption{Visual representation of connections within both empirical and predicted FCs found significant for classifying into SZ and HC. Connections within the same brain network  were visually highlighted with distinct colors, while connections across different networks were depicted in gray. The thickness of the edges reflects the strength of their respective connections in the obtained explanation graph.} \label{fig2}
\vspace{-10 pt}
\end{figure}
\vspace{-5 pt}
\subsection{Qualitative Evaluation}
The ROIs within brain networks can be categorized into neural systems based on their structural and functional roles, defined by a specific parcellation atlas. This categorization helps in understanding the interpretations provided from a neuroscience perspective. In this work, we assigned the ROI nodes from each dataset to seven well-known neural systems using the NeuroMark atlas  \cite{21}. These systems include the subcortical network (SCN), auditory network (ADN), sensorimotor network (SMN), visual network (VSN), cognitive control network (CON), default mode network (DMN), and cerebellum network (CBN).

To provide interpretation and clinical validation of our approach, we employed GNNExplainer \cite{25} in our work to identify the connections in both empirical and predicted FC that significantly contributed towards SZ or HC classification. As depicted in Fig. \ref{fig2}, denser connections were observed within the functional brain networks of HC compared to individuals with SZ. Notably, both empirical and predicted FC revealed significant interactions within DMN and CON for SZ classification. These findings align with previous studies \cite{L_9,5,4,6} and underscore the clinical observation that SZ perturbs the human brain's structural-functional dynamic states \cite{L_10,L_9,L_11}. Moreover, related overlapping connections were observed significant for SZ classification employing both empirical and predicted FC which also validated our predicted FC from a neuroscience perspective.
\vspace{-5 pt}
\section{Conclusions}
\vspace{-5 pt}
We proposed a novel approach that couples physics model with deep learning to predict FC from SC by exploring the neuroscience theory suggesting that random functional fluctuations emerge from a dynamic network of neural oscillations. The predicted FCs were then utilized with SCs in a multi-modal configuration by incorporating correlation-based fusion along with the final classification task simultaneously in an end-to-end fashion. Our introduced framework not only tackles the challenge of functional data scarcity but also improves SZ classification concurrently. The robustness of our method was validated through thorough experimentation on the FBIRN dataset, yielding significant performance improvements against several SOTA baselines alongside providing necessary explanations.

\bibliographystyle{splncs04}
\bibliography{ref}
\end{document}